\documentclass[aps,pra,twocolumn,amsmath,amssymb]{revtex4}
\usepackage{graphicx}
\usepackage{amsmath}

\begin{document}

\title{Highly Polarized Fermi Gases across a Narrow Feshbach Resonance}
\author{Ran Qi and Hui Zhai}
\affiliation{Institute for Advanced Study, Tsinghua University, Beijing, 100084, China}
\date{\today}
\begin{abstract}

We address the phase of a highly polarized Fermi gas across a {\it narrow} Feshbach resonance starting from the problem of a single down spin fermion immersed in a Fermi sea of up spins. Both polaron and pairing states are considered using the variational wave function approach, and we find that the polaron to pairing transition will take place at the BCS side of the resonance, strongly in contrast to a wide resonance where the transition is located at the BEC side. For pairing phase, we find out the critical strength of repulsive interaction between pairs above which the mixture of pairs and fermions will not phase separate. Therefore, nearby a narrow resonance, it is quite likely that magnetism can coexist with $s$-wave BCS superfluidity at large Zeeman field, which is a remarkable property absent in conventional BCS superconductors (or fermion pair superfluids).
\end{abstract}
\maketitle

Whether an $s$-wave superconductor (or fermion pair superfluid) can coexist with magnetism is a long-standing issue in condensed matter physics. Back to 1960s', Chandrasekhar and Clogston independently considered the response of a BCS superconductor to spin polarization due to a Zeeman field \cite{Clogston, Chandrasekhar}. They found that an $s$-wave superconductor will remain unpolarized until a critical Zeeman energy $h_{\text{c}}\sim\Delta/\sqrt{2}$ with $\Delta$ being the pairing gap, at which the system undergoes a sharp phase transition to a partially polarized normal state, and this critical field is now known as Chandrasekhar-Clogston (CC) limit of superconductor. In this scenario,  superconductivity can not coexist with magnetism. Later on, there are several proposals for magnetized $s$-wave superconducting states, and the most famous ones are the Fulde-Ferrell-Larkin-Ovchinnikov state \cite{FFLO} and the Sarma State \cite{Sarma}. However, so far none of them have been firmly observed in an $s$-wave BCS superconductors.

In the last few years, this problem has been revisited by a series of experiments on two-component Fermi gases with population imbalance \cite{Ketterle,Ketterle_mix,Randy,Randy2,Salamon}.
Experiments have reached a consensus that at resonance regime and the BCS side, there is a CC limit, where a direct transition from a fully paired fermion superfluid to a partially polarized normal state has been observed, and no evidence of a magnetized superfluid has been found \cite{Ketterle,Randy2,Salamon,review}.
However, all these studies were done across a wide resonance. Recently, several experimental groups have begun to study narrow Feshbach resonances, such as $^6$Li at 543.25G \cite{Li} or $^6$Li-$^{40}$K mixture at 154.719G \cite{LiK}, where the effect of a finite resonance width needs to be taken into account. In this letter we find that the resonance width indeed has dramatic effects on physics of highly polarized Fermi gases.

In contrast to a wide resonance, to character a Fermi gas nearby a narrow resonance one not only needs $k_{\text{F}}a^0_{\text{s}}$, where $a^0_{\text{s}}$ is zero-energy scattering length between fermions, but one also needs to consider $\hbar^2k_{\text{F}}/(2mWa_{\text{bg}})$ with $m$ being atom mass $W$ the resonance width and $a_{\text{bg}}$ the background scattering length. If $\hbar^2k_{\text{F}}/(2mWa_{\text{bg}})\gtrsim 1$ the resonance is considered to be a narrow one. Another important quantity is $k_{\text{F}}a_{\text{bb}}$ with $a_{\text{bb}}$ being the scattering length between closed channel molecules, the effect of this quantity will be analyzed later. In this work we focus on highly polarized limit and show all above three parameters play an important role in determining the nature of many-body phases. The studies of this work contain two parts:

Firstly, we consider a single down spin immersed in a Fermi sea of up spins. Two different types of states are compared, which are polaronic state and pairing state. For polaronic state, the single down spin is dressed by particle-hole pairs of up spins, and becomes a fermionic quasi-particle \cite{polaron}. If this state has lower energy, the system will be a normal state of polaron Fermi liquid at sufficient high polarization. For pairing state, one of the up spins will form a bound state with the single down spin. If this state has lower energy, each down spin will form a pair, and the system will be a mixture of condensed pairs and majority fermions. For wide resonance, a polaron to pairing transition takes place at the BEC side of the resonance \cite{polaron-molecule,DMC,PM_exp}. Here we show that as the width of resonance gets narrower, the transition point will be shifted toward the BCS side. We find out how the critical value of $(k_{\text{F}}a^0_{\text{s}})_{\text{c}}$ changes with the quantity $\hbar^2k_{\text{F}}/(2mWa_{\text{bg}})$.

Secondly, when the pairing state has lower energy, the mixture of pairs and fermions may phase separate due to the repulsion between pairs and fermions. A strong enough repulsion between pairs are crucial to stabilize a uniform mixture. For a given $\hbar^2k_{\text{F}}/(2mWa_{\text{bg}})$, we find out the critical repulsion $(k_{\text{F}}a_{\text{bb}})_{\text{c}}$ as a function of $k_{\text{F}}a^0_{\text{s}}$.

Hence, we conclude that when $1/(k_{\text{F}}a^0_{\text{s}})>1/(k_{\text{F}}a^0_{\text{s}})_{\text{c}}$ and $k_{\text{F}}a_{\text{bb}}>(k_{\text{F}}a_{\text{bb}})_{\text{c}}$, it is energetically favorable for minority fermions to form pairs, and condensate of fermion pairs can uniformly mix with majority fermions, that is to say, magnetism can coexist with fermion pair superfluids at highly polarized Fermi gases. The fact that this can happen at the BCS side and resonance regime represents a significant distinction between narrow and wide resonances.  As far as the response to spin polarization is concerned, at resonance, or even at the BCS side, this system behaves similar as the BEC side of a wide resonance. This picture is also consistent with a recent high-temperature study of narrow resonance \cite{cui}.

{\it Model:}  We use the following two-channel model $\hat{H}=\hat{H}_0+\hat{V}_\text{c}+\hat{V}_{\text{bg}}+\hat{V}_{\text{bb}}$ to describe a narrow resonance
\begin{align}
\hat{H}_0&=\sum_{\mathbf{k}}(\epsilon_{\mathbf{k}}^{\text{b}}+\nu_0)b^{\dag}_{\mathbf{k}}b_{\mathbf{k}}
+\sum_{\mathbf{k}}(\epsilon^{\text{u}}_{\mathbf{k}}u^{\dag}_{\mathbf{k}}u_{\mathbf{k}}+\epsilon^{\text{d}}_{\mathbf{k}}d^{\dag}_{\mathbf{k}}d_{\mathbf{k}}) \label{H0} \\
\hat{V}_{\text{c}}&=g_0\sum_{\mathbf{k}\mathbf{q}}\Lambda_{\mathbf{k}}(b^{\dag}_{\mathbf{q}}d_{\mathbf{q-k}}u_{\mathbf{k}}+u^{\dag}_{\mathbf{k}}d^{\dag}_{\mathbf{q-k}}b_{\mathbf{q}}) \label{Vc}\\
\hat{V}_{\text{bg}}&=U_0\sum_{\mathbf{k}\mathbf{k^\prime}\mathbf{q}}\Lambda_{\mathbf{k}}\Lambda_{\mathbf{k^\prime}}u^{\dag}_{\mathbf{k}}d^{\dag}_{\mathbf{q-k}}d_{\mathbf{q-k^\prime}}u_{\mathbf{k^\prime}}\label{Vbg}\\
\hat{V}_{\text{bb}}&=\frac{1}{2}U_{\text{bb}}\sum\limits_{\mathbf{k}\mathbf{k^\prime}\mathbf{q}}
b^{\dag}_{\mathbf{k}}b^{\dag}_{\mathbf{q-k}}b_{\mathbf{q-k^\prime}}b_{\mathbf{k^\prime}}\label{Vbb}
\end{align}
where $u^\dag$ and $d^\dag$ are creation operators for majority up spin and minority down spin, respectively, and $b^\dag$ is creation operator for bosonic closed channel molecule. $\epsilon_{\mathbf{k}}^{\text{b}}=\hbar^2 {\bf k}^2/[2(m^{\text{u}}+m^{\text{d}})]$, and $\epsilon_{\mathbf{k}}^{\text{u/d}}=\hbar^2 {\bf k}^2/(2m^{\text{u/d}})$. $\gamma=m^{\text{d}}/m^{\text{u}}$ is the mass ratio. $\hat{V}_{\text{c}}$ and $\hat{V}_{\text{bg}}$ represent the interchannel coupling and the background scattering, respectively. $\Lambda_{\mathbf{k}}=\Theta(\Lambda-|{\bf k}|)$ and $\Lambda$ is the momentum cutoff. In the Hamiltonian, the molecule detuning $\nu_0$, the inter-channel coupling $g_0$ and the background interaction parameter $U_0$ are bare quantities with $\Lambda$ dependence, which need to be renormalized as follows \cite{renormalization,bruun}: $\nu_0(\Lambda)=\nu_\text{r}-[1-Z(\Lambda)]g^2_\text{r}/U_\text{r}$, $g_0(\Lambda)=Z(\Lambda)g_{\text{r}}$, and $U_0(\Lambda)=Z(\Lambda)U_\text{r}$, where $Z(\Lambda)=(1-U_\text{r}m_\text{r}\Lambda/\pi^2)^{-1}$, $1/m_{\text{r}}=1/m^{\text{u}}+1/m^{\text{d}}$ and $U_{\text{r}}=2\pi a_{\text{bg}}/m_{\text{r}}$. The renormalized quantities $U_\text{r}$, $g_\text{r}$ and $\nu_\text{r}$ are related to $a_\text{s}$ as
\begin{equation}
\frac{2\pi a_{\text{s}}(E) }{m_{\text{r}}}\!=\!\left[\left(U_0+\frac{g^2_0}{E-\nu_0}\right)^{-1}\!+\!\frac{m_{\text{r}}\Lambda}{\pi^2}\right]^{-1}\!=\!U_{\text{r}}\!+\!\frac{g^2_{\text{r}}}{E-\nu_{\text{r}}}\nonumber
\end{equation}
and the zero-energy scattering length $a^0_{\text{s}}$ is given by $a^0_{\text{s}}=m_\text{r}(U_\text{r}-g^2_{\text{r}}/\nu_\text{r})/(2\pi)$. Denoting $\nu_{\text{r}}=\Delta\mu(B-B_0)$, where $B_0$ is the location of the resonance and $\Delta\mu$ is the difference of magnetic moment between two channels, and introducing $W=g^2_{\text{r}}/U_{\text{r}}$, we have $a^0_{\text{s}}=a_{\text{bg}}\{1-W/[\Delta\mu(B-B_0)]\}$, and $a_{\text{s}}(E)=a_{\text{bg}}\{1+W/[E-\Delta\mu(B-B_0)]\}$ where $Wa_{\text{bg}}$ is always positive. For $U_{\text{bb}}$, since we only consider it to the mean-field order, we will take it as $U_{\text{bb}}=4\pi\hbar^2 a_{\text{bb}}/(m^\text{u}+m^\text{d})$. Our following results will be presented in terms of physical parameters ($W$, $B_0$, $k_{\text{F}}a^0_{\text{s}}$, $k_{\text{F}}a_{\text{bg}}$ and $k_{\text{F}}a_{\text{bb}}$).

{\it Polaronic State:} We first adopt the following variational wave function which includes one particle-hole contribution
\begin{eqnarray}
|\psi^\text{p}\rangle\!=\!\Big[\phi_0 d^{\dag}_0\!+\!\sum^{\prime}_{\mathbf{kq}}\phi_{\mathbf{kq}}u^{\dag}_{\mathbf{k}} d^{\dag}_{\mathbf{q-k}}u_{\mathbf{q}}\!+\!\sum^{\prime}_{\mathbf{q}}\eta_{\mathbf{q}}b^{\dag}_{\mathbf{q}}u_{\mathbf{q}}\Big]|\text{FS}\rangle.\label{polaron ansatz}
\end{eqnarray}
Here and below, all the summations with the prime ($^\prime$) of $\mathbf{k}$ and $\mathbf{q}$ are restricted to $|{\bf k}|>k_{\text{F}}$ and $|{\bf q}|<k_{\text{F}}$, respectively.
After energy minimization we obtain a self-consistent equation for polaron energy
\begin{eqnarray}
E=\sum_{{\bf q}}^\prime\Gamma_2(\mathbf{q},E+\epsilon^\text{u}_{\mathbf{q}}),\label{polaron_E}
\end{eqnarray}
where $\Gamma_2({\bf q},E+\epsilon^\text{u}_{\mathbf{q}})$ coincides with two-particle vertex with total momentum $\mathbf{q}$ and total energy $E+\epsilon^\text{u}_{\mathbf{q}}$ within the ladder approximation. This is because the variational wave function $|\psi^\text{p}\rangle$ describes the processes that an up spin (taken out from an occupied state ${\bf q}$) and the single down spin undergo repeated scattering, as well as coherent conversion between open and closed channels. This physical process is precisely what is captured by the ladder approximation.

\begin{figure}[tbp]
\includegraphics[height=1.6in, width=3.4in]
{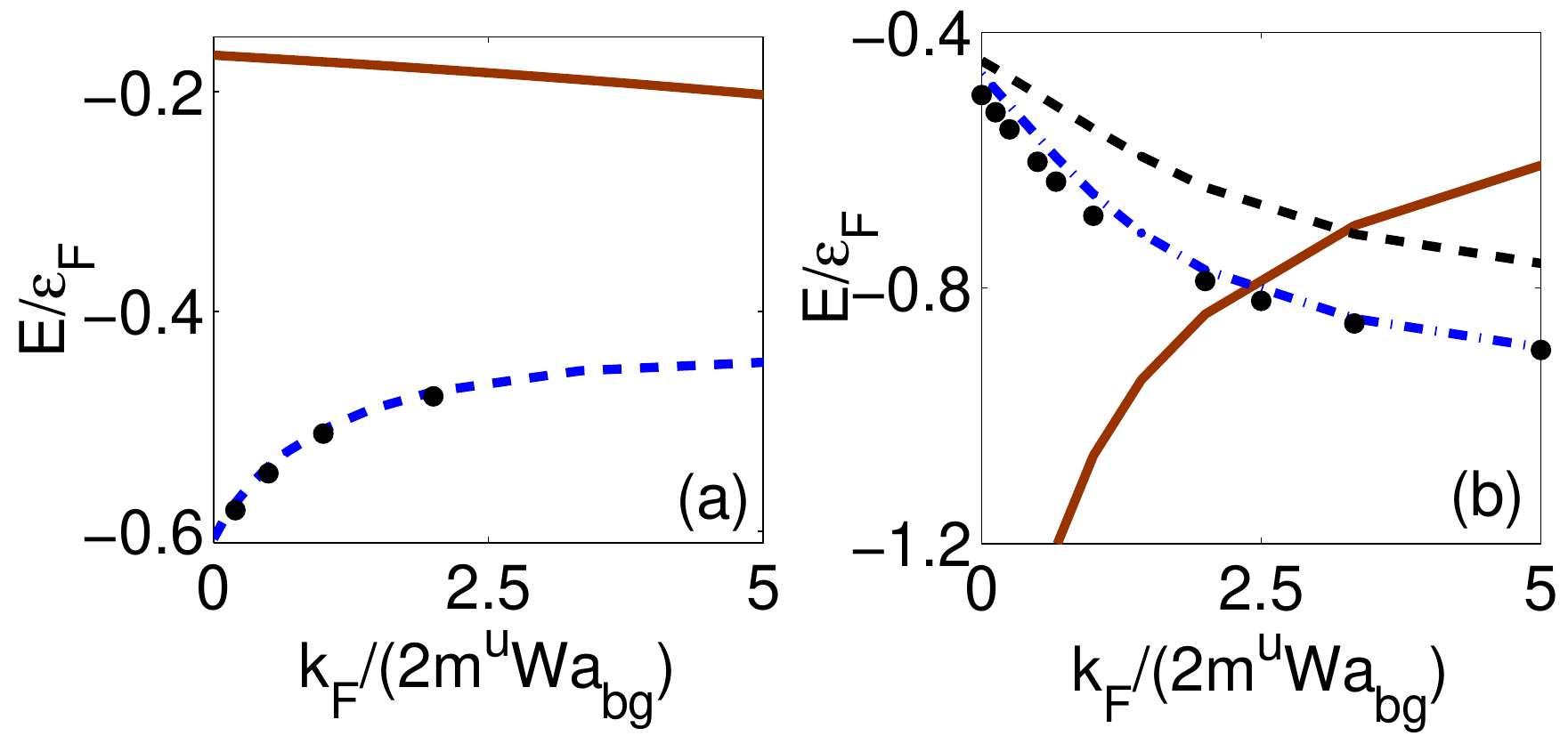}
\caption{(Color online) Polaron energy as a function of  $\hbar^2k_{\text{F}}/(2m^uWa_{\text{bg}})$. (a) $\gamma=1$ and different interaction parameters. $1/(k_{\text{F}}a^0_{\text{s}})=0$ (dash line) and $1/(k_{\text{F}}a^0_{\text{s}})=-2$ (solid line). (b) $1/(k_{\text{F}}a^0_{\text{s}})=0$ with different mass ratio. $\gamma=6/40$ (solid line), $\gamma=40/6$ (dash line) and $\gamma=\infty$ (dash-dotted line). All curves are results from one particle-hole approximation, while solid black dots show results including two particle-hole pairs contributions. $k_{\text{F}}a_{\text{bg}}$ is chosen as $-0.1$. We will also set $\hbar=1$ in all figures to simplify the presentation. \label{polaron}}
\end{figure}

The explicit form of $\Gamma_2({\bf q},E+\epsilon^\text{u}_{\mathbf{q}})$ is given as
\begin{equation}
\Gamma_2({\bf q},E\!+\!\epsilon^\text{u}_{\mathbf{q}})\!=\!\left[\!\frac{m_{\text{r}}}{2\pi a_{\text{s}}(E\!+\!\epsilon^\text{u}_{\mathbf{q}}\!-\!\epsilon^\text{b}_{\bf q})}\!+\!I({\bf q},E\!+\!\epsilon^\text{u}_{\mathbf{q}})\!\right]^{-1}, \label{self-consistent}
\end{equation}
where $I({\bf q},E+\epsilon^\text{u}_{\mathbf{q}})=\sum^\prime_{\mathbf{k}}1/[\epsilon^{\text{u}}_{\mathbf{k}}+\epsilon^\text{d}_{\mathbf{q-k}}-(E+\epsilon^\text{u}_{\mathbf{q}})]
-\sum_{\mathbf{k}}1/(\epsilon^\text{u}_{\mathbf{k}}+\epsilon^\text{d}_{\mathbf{-k}})$.
The difference between Eq. (\ref{self-consistent}) and the previous results for a wide resonance \cite{polaron,polaron-molecule} is that a constant $a_{\text{s}}$ is replaced by an energy dependent one $a_{\text{s}}(E+\epsilon^\text{u}_{\mathbf{q}}-\epsilon^\text{b}_{\bf q})$, where $E+\epsilon^\text{u}_{\mathbf{q}}-\epsilon^\text{b}_{\bf q}$ represents the energy of the relative motion for two atoms undergoing repeated scatterings.

The polaron energy as a function of $\hbar^2k_{\text{F}}/(2m^uWa_{\text{bg}})$ is plotted in Fig. (\ref{polaron}) from solving self-consistency equation Eq. (\ref{polaron_E}). As one can see, when the dimensionless parameter $\hbar^2k_{\text{F}}/(2m^uWa_{\text{bg}})$ increases from zero, (i) for $\gamma=1$, and nearby resonance $1/(k_{\text{F}}a^0_{\text{s}})\approx 0$ polaron energy $E$ will increase, while at the BCS side $1/(k_{\text{F}}a^0_{\text{s}})\ll 0$, $E$ will decrease; and (ii) at resonance, if $\gamma$ is greater than a critical value, $E$ will also decrease. We have also checked the energy convergence by considering two particle-hole contributions \cite{Combescot2,supple}. The numerical solutions with two particle-hole contributions are shown as the dots in Fig. (\ref{polaron}). For $\gamma=1$, one can see that the corrections from two particle-hole pairs are always negligibly small, and it becomes even smaller as $|W|$ decreases. While for $\gamma\rightarrow\infty$, the deviation is a little larger (dashed line and the dots in (b)), as already noted for a wide resonance in \cite{Combescot2}, but it is still within only a few percents. This result justifies the validity of the expansion in terms of the number of particle-hole pairs in computing energy for a narrow resonance.

\begin{figure}[tbp]
\includegraphics[height=1.6in, width=3.4in]
{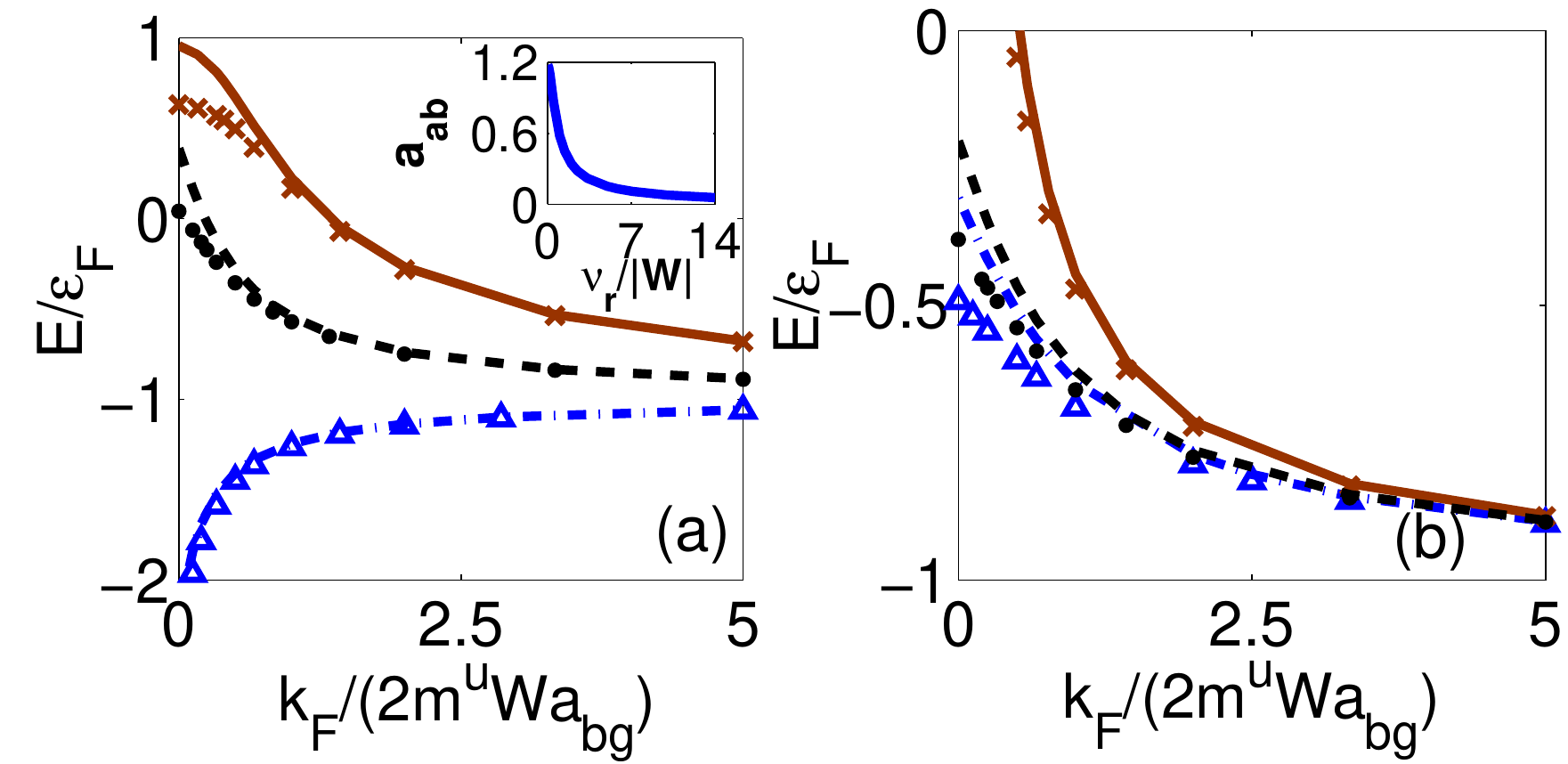}
\caption{(Color online) Pairing state energy as a function of $\hbar^2k_{\text{F}}/(2m^uWa_{\text{bg}})$. (a) mass ratio $\gamma=1$ and different interaction parameters.  $1/(k_{\text{F}}a^0_{\text{s}})=0$ (dash line), $1/(k_{\text{F}}a^0_{\text{s}})=-1$ (solid line) and $1/(k_{\text{F}}a^0_{\text{s}})=1$ (dash-dotted line). (b) $1/(k_{\text{F}}a^0_{\text{s}})=0$ but different mass ratio. $\gamma=6/40$ (solid line), $\gamma=40/6$ (dash line) and $\gamma=\infty$ (dash-dotted line). All the curves are computed from bare pairs, while the dots, crosses and triangles are results with one particle-hole contributions. Inset of (a): the atom-dimer scattering length $a_{\text{ab}}$ (in unit of $\sqrt{m^uE_{\text{b}}}/\hbar$ with $E_{\text{b}}$ the two-body binding energy) as a function of $\nu_{\text{r}}/W$. $k_{\text{F}}a_{\text{bg}}$ is chosen as $-0.1$. \label{molecule}}
\end{figure}

{\it Pairing State:} For pairing state, we use the variational wave function first introduced in Ref. \cite{polaron-molecule}:
\begin{eqnarray}
|\psi^\text{m}\rangle&=&\Big[\eta_0 b_0^{\dag}+\sum'_{\mathbf{k}}A_{\mathbf{k}}u^{\dag}_{\mathbf{k}}d^{\dag}_{-\mathbf{k}}+\sum'_{\mathbf{k}\mathbf{q}}\phi_{\mathbf{k}\mathbf{q}}b^{\dag}_{\mathbf{q-k}}u^{\dag}_{\mathbf{k}}u_{\mathbf{q}}\nonumber\\
&~&+\sum'_{\mathbf{k'}\mathbf{k}\mathbf{q}}\Phi_{\mathbf{k}\mathbf{k'}\mathbf{q}}u^{\dag}_{\mathbf{k'}}d^{\dag}_{\mathbf{q-k-k'}}u^{\dag}_{\mathbf{k}}u_{\mathbf{q}}\Big]|\text{FS}'\rangle
\end{eqnarray}
where $|\text{FS}'\rangle$ refers to the Fermi sea with one spin-$\uparrow$ particle removed from the Fermi surface of $|\text{FS}\rangle$ in the polaron state (\ref{polaron ansatz}).
If we only consider bare pair wave function without including particle-hole contribution, the pairing state energy is given by $\Gamma^{-1}_2(0,E+\epsilon_F)=0$, as shown in the lines of Fig. (\ref{molecule}). The interaction between pair and majority up spins can be described by including particle-hole pairs. Up to one particle-hole pair, by minimizing energy we obtain a closed integral equation \cite{supple}, and
 the numerical solution of these equations are also shown in Fig. (\ref{molecule}).
 We find that as $k_{\text{F}}/(2m^uWa_{\text{bg}})$ increases, the pairing state energy decreases at the BCS side and at resonance regime,
 despite of different mass ratios; while it increases at the BEC side.
 Another important feature one can find from Fig. (\ref{molecule}) is that in the limit $W\rightarrow 0$, the pairing state energy always saturates to $-\epsilon_F$. This can be understood as follows: when one down spin is added into the system, an up spin is taken out from the Fermi sea (subtract energy $\epsilon_F$) to form a pair with the down spin, whose energy approaches $\nu_{\text{r}}$ in the limit $W\rightarrow 0$. Thus the pairing state energy should approach $-\epsilon_F+\nu_{\text{r}}$ \cite{Leo}. Moreover, at any fixed $a^0_{\text{s}}$ the ratio $W/\nu_{\text{r}}$ is fixed, thus $\nu_{\text{r}}\rightarrow 0$ and therefore pairing state energy always approaches $-\epsilon_F$ as  $W\rightarrow 0$. This also indicates that the interaction between a pair and the residual majority atoms vanishes in the limit $W\rightarrow 0$. We have performed a three-body calculation and find out the atom-dimer scattering length $a_{\text{ab}}$ from the asymptotic behavior of the three-body wave function \cite{petrov,supple}, as shown in the inset of Fig. (\ref{molecule})(a), which indeed shows $a_{\text{ab}}\rightarrow 0$ as $W\rightarrow 0$.

\begin{figure}[tbp]
\includegraphics[height=2.7in, width=3.4in]
{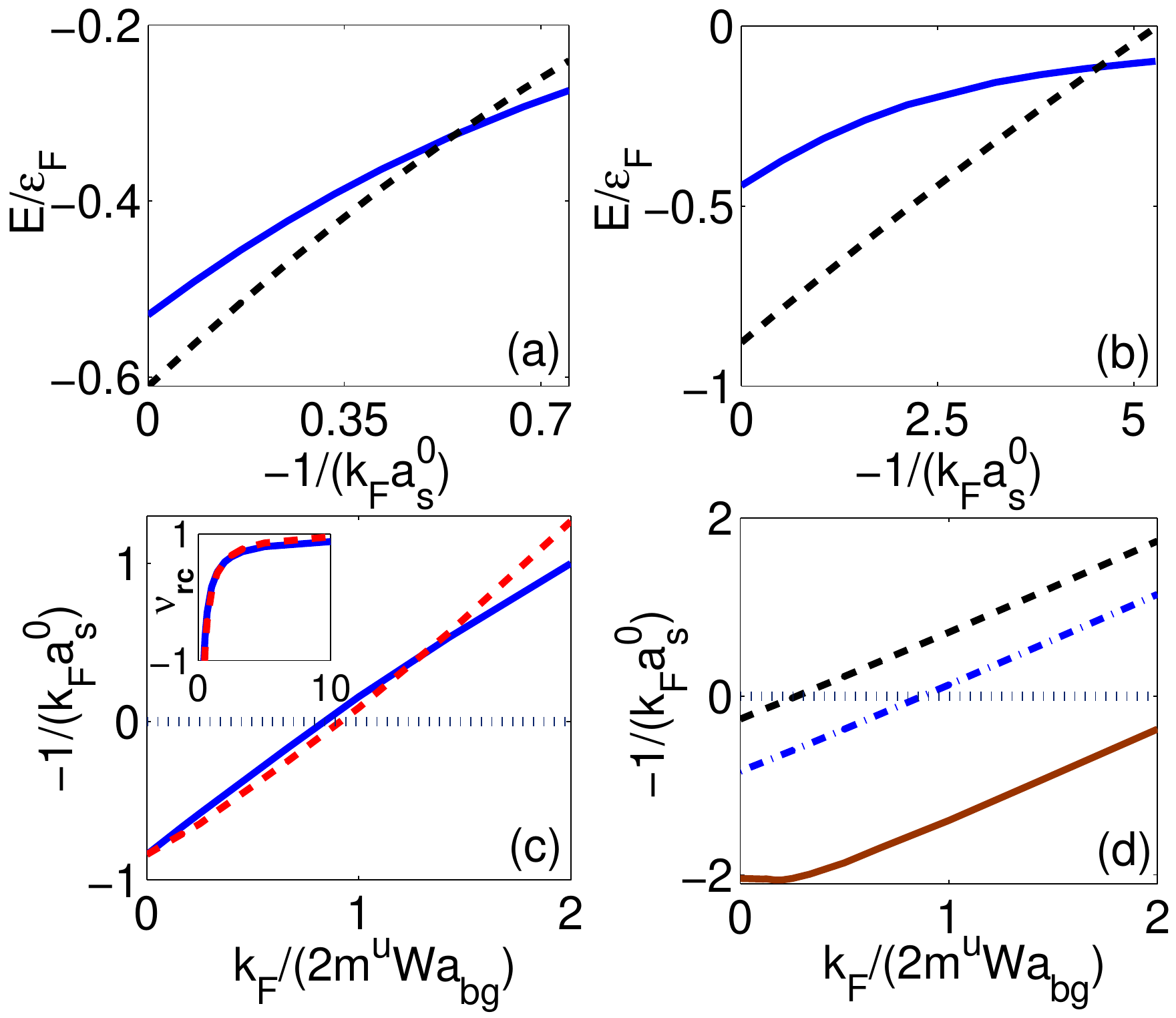}
\caption{(Color online) Comparison of polaron energy (solid line) and pairing state energy (dash line) as a function of zero-energy scattering length for $^6$Li-$^{40}$K mixture (a) and $^6$Li (b). For (a), we take $k_{\text{F}}a_{\text{bg}}=0.022$ and $W=54.91\epsilon_F$ \cite{Grimm}, and for (b) we take $k_{\text{F}}a_{\text{bg}}=0.016$ and $W=12.2\epsilon_F$ \cite{Li} . In (a), $^{40}$K is taken as minority component. Panels (c) and (d) show critical value of the transition $1/(k_{\text{F}}a^0_{\text{s}})_{\text{c}}$ as a function of $\hbar^2k_{\text{F}}/(2m^uWa_{\text{bg}})$. For (c), $\gamma=1$ but different $k_{\text{F}}a_{\text{bg}}$. $k_{\text{F}}a_{\text{bg}}=-0.1$ (solid line) and $k_{\text{F}}a_{\text{bg}}=0.1$ (dashed line).  For (d), $k_{\text{F}}a_{\text{bg}}=0.01$ but different $\gamma$. $\gamma=6/40$ (solid line), $\gamma=1$ (dash-dotted line) and $\gamma=40/6$ (dash line). Inset of (c), the critical value of transition in term of $\nu^{\text{c}}_{\text{r}}$ (in unit of $\epsilon_F$) as a function of $\hbar^2k_{\text{F}}/(2m^uWa_{\text{bg}})$.  \label{p-m}}
\end{figure}

{\it Polaron-Pairng Transition:} The transition from polaronic state to pairing state can now be determined by comparing their energies. In Fig. (\ref{p-m})(a-b) we consider two concrete samples studied in current experiments: $^6$Li at 543.25G and $^6$Li-$^{40}$K mixture at 154.719 G, and the parameters are typical values taken from experimental papers \cite{Li,LiK,Grimm}. We found that in both cases, the polaron to pairing transition is located at the BCS side of the resonance, which is away from resonance with $\Delta\mu(B-B_0)$ on the order of $\epsilon_F$. At the transition points, $1/(k_{\text{F}}a^0_{\text{s}})=-4.35$ for $^6$Li and $1/(k_{\text{F}}a^0_{\text{s}})=-0.55$ for $^6$Li-$^{40}$K mixture, where the systems are very BCS-like. This transition has also been observed in a recent experiment on $^6$Li-$^{40}$K mixture and the transition is indeed observed at the BCS side \cite{Grimm}.

In Fig. \ref{p-m}(c-d), we plot the critical value $1/(k_{\text{F}}a^0_{\text{s}})$ for polaron to pairing transition as a function of $k_{\text{F}}/(2m^uWa_{\text{bg}})$. One finds that when $\hbar^2k_{\text{F}}/(2m^uWa_{\text{bg}})\gtrsim 1$, the transition will be shifted to the BCS side. This condition is equivalent to $|W|/\epsilon_F\lesssim 1/|k_{\text{F}}a_{\text{bg}}|$. Since usually $k_{\text{F}}a_{\text{bg}}\ll 1$, it means that the resonance width does not need to be very narrow. One also notes that the transition point is not sensitive to the value of $k_{\text{F}}a_{\text{bg}}$ itself (Fig. (\ref{p-m})(c)), but is sensitive to mass ratio (Fig. \ref{p-m}(d)). The inset of Fig. (\ref{p-m})(c) shows that in the limit $W\rightarrow 0$, the critical point will approach $\nu^{\text{c}}_{\text{r}}\rightarrow \epsilon_F$, which means that the pairing state will be favored once the energy of closed channel molecule is below the Fermi energy.

\begin{figure}[tbp]
\includegraphics[height=1.6in, width=3.3in]
{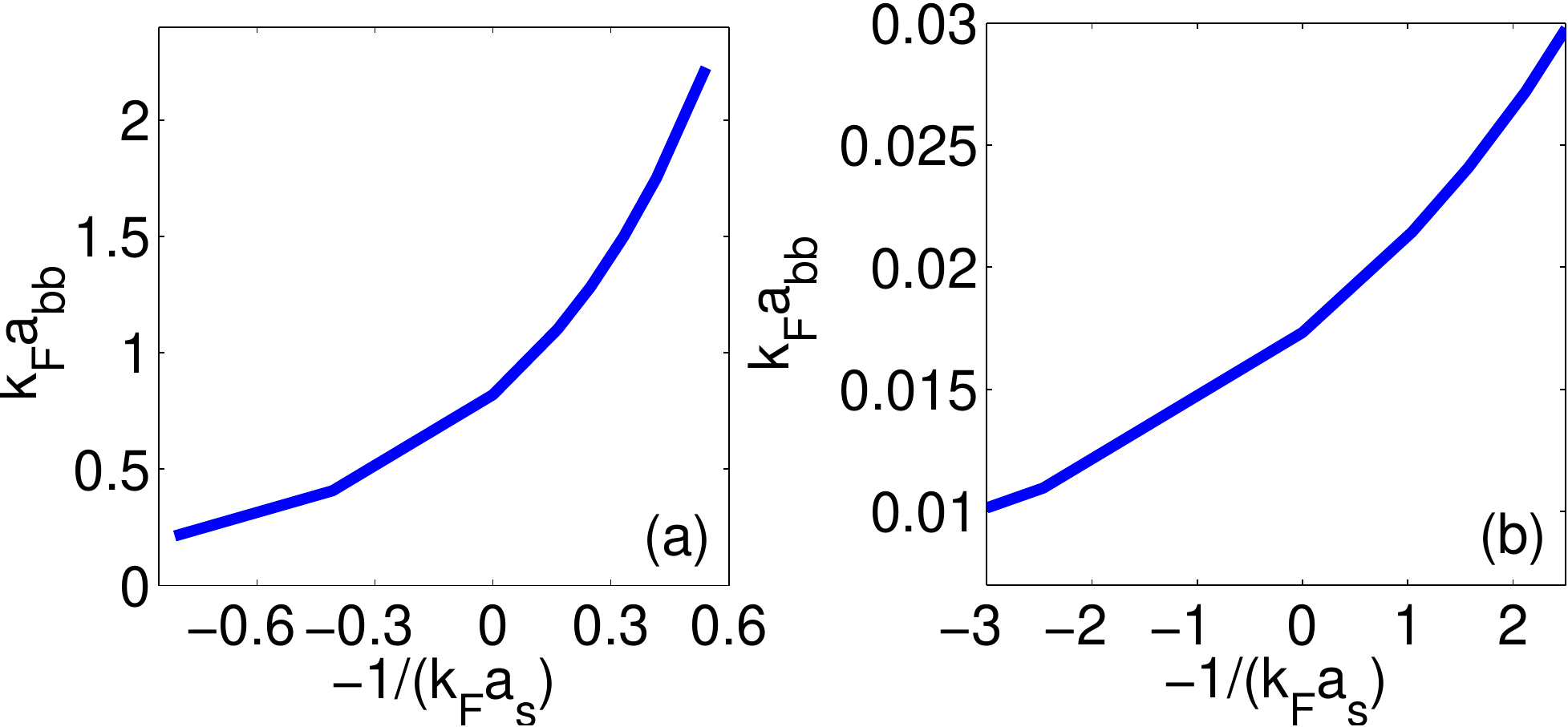}
\caption{(Color online) Critical value required for $k_{\text{F}}a_{\text{bb}}$ to prevent phase separation as a function of $1/(k_{\text{F}}a^0_{\text{s}})$, for $^6$Li-$^{40}$K mixture (a) and $^6$Li (b).  \label{stability}}
\end{figure}

{\it Stability of the Mixture:} Discussion above concludes that for sufficiently narrow resonance, the highly polarized Fermi gases contain a mixture of majority fermions and bosonic pairs. Next question is whether they will uniformly mix or phase separate. To answer this question, we note that for very low density of down spins and sufficiently narrow resonance, we can expand the equation-of-state in terms of the density of bosonic pairs $n_{\text{b}}$ up to the second order \cite{stability}
\begin{eqnarray}
\mathcal{E}=\mathcal{E}_F+\mu^0_\text{b}n_\text{b}+\frac{1}{2}g n_\text{b}^2\label{energy}
\end{eqnarray}
where $\mathcal{E}_F$ is the energy density of spin-$\uparrow$ Fermi sea, $\mu_\text{b}^0=E+\epsilon_F$ where $E$ is the pairing state energy computed above. The repulsion $g=\frac{4\pi \hbar^2}{m^\text{u}+m^\text{d}}a_{\text{bb}}+g_{\text{ind}}$ contains the contribution from the bare interaction between closed channel molecules and the induced interaction $g_{\text{ind}}$ from the inter-channel coupling, which is calculated within Born approximation \cite{supple}. From Eq. (\ref{energy}) we obtain:
\begin{align}
\mu_\text{b}&=\frac{\partial \mathcal{E}}{\partial n_b}=\mu^0_\text{b}+gn_\text{b}\\
\mu_{\uparrow}&=\frac{\partial \mathcal{E}}{\partial n_{\uparrow}}=\epsilon_F+\frac{\partial \mu^0_\text{b}}{\partial n_{\uparrow}}n_\text{b}+O(n_\text{b}^2)
\end{align}
The stability condition against phase separation is given by $\frac{\partial \mu_{\uparrow}}{\partial n_{\uparrow}}\frac{\partial \mu_\text{b}}{\partial n_\text{b}}-\frac{\partial \mu_{\uparrow}}{\partial n_\text{b}}\frac{\partial \mu_\text{b}}{\partial n_{\uparrow}}>0$ \cite{Pethick}, from which we can determine the critical value for $a_{\text{bb}}$. The results are plotted in Fig. \ref{stability} for $^6$Li-$^{40}$K mixture (a) and  $^6$Li (b) in the regime where pairing state is favorable. We can see that it requires $k_\text{F}a_{\text{bb}}>0.81$ for $^6$Li-$^{40}$K mixture, and $k_\text{F}a_{\text{bb}}>0.017$ for $^6$Li at resonance. Very likely, this condition can be satisfied in $^6$Li but not in $^6$Li-$^{40}$K mixture.

{\it Acknowledgements.} We thank Xiaoling Cui and Zeng-Qiang Yu for helpful discussions. This work is supported by Tsinghua University Initiative Scientific Research Program. HZ is supported by NSFC under Grant No. 11004118 and No. 11174176, and NKBRSFC under Grant No. 2011CB921500. RQ is supported by NSFC under Grant No. 11104157.

{\bf Note Added:} Recently, we became aware of the experimental work from Innsbruck group in which the polaron properties and the polaron to pairing transition has been observed \cite{Grimm}. We also became aware of two other related theoretical works \cite{related1,related2}.

\newpage

\begin{widetext}

\section{Supplementary Material}

\subsection{Polaron energy including two particle-hole contributions}
In this section of supplementary material, we derive the self-consistent equation for polaron energy including two particle-hole excitations.

The Hamiltonian under consideration is given as equation (1)-(3) in our paper, and the variational wave function is written as:
\begin{eqnarray}
|\psi\rangle&=&|\hat{\phi}_0\rangle+|\hat{\phi}_{\mathbf{kq}}\rangle+|\hat{\eta}_{\mathbf{q}}\rangle+|\hat{\alpha}\rangle+|\hat{\beta}\rangle\\
|\hat{\phi}_0\rangle&=&\phi_0 d^{\dag}_{0}|FS\rangle ,~~~~~~~~~~~|\hat{\phi}_{\mathbf{kq}}\rangle=\sum'_{\mathbf{kq}}\phi_{\mathbf{kq}}u^{\dag}_{\mathbf{k}} d^{\dag}_{\mathbf{q-k}}u_{\mathbf{q}}|FS\rangle ,~~~~~~~~~~~|\hat{\eta}_{\mathbf{q}}\rangle=\sum'_{\mathbf{q}}\eta_{\mathbf{q}}b^{\dag}_{\mathbf{q}}u_{\mathbf{q}}|FS\rangle\\
|\hat{\alpha}\rangle&=&\sum'_{\mathbf{kk'qq'}}\alpha_{\mathbf{kk'qq'}}u^{\dag}_{\mathbf{k}}u^{\dag}_{\mathbf{k'}}d^{\dag}_{\mathbf{q+q'-k-k'}}u_{\mathbf{q}}u_{\mathbf{q'}}|FS\rangle ,~~~~~~~~~~~~
|\hat{\beta}\rangle=\sum'_{\mathbf{kqq'}}\beta_{\mathbf{kqq'}}b^{\dag}_{\mathbf{q+q'-k}}u^{\dag}_{\mathbf{k}}u_{\mathbf{q}}u_{\mathbf{q'}}|FS\rangle
\end{eqnarray}
where $|FS\rangle$ is the Fermi sea of N spin-$\uparrow$ particles, $|\hat{\phi}_{\mathbf{kq}}\rangle$ and $|\hat{\eta}_{\mathbf{q}}\rangle$ correspond to single particle-hole excitation while $|\hat{\alpha}\rangle,~|\hat{\beta}\rangle$ represents the contribution from two particle-hole excitation. All momentum summations are restricted within $|{\bf q}|<k_\text{F},~|{\bf k}|>k_\text{F}$ as mentioned in our paper.
Now we solve the Schr\"odinger equation $\hat{H}|\psi\rangle=E|\psi\rangle$ and drop all the terms beyond two particle-hole excitation, we get the following set of equations:
\begin{eqnarray}
E\phi_0&=&\sum'_{\mathbf{q}}\varphi_{\mathbf{q}}\label{p1}\\
E_{\mathbf{kq}}\phi_{\mathbf{kq}}&=&-g_0\eta_{\mathbf{q}}-U_0\phi_0-\varphi_{\mathbf{q}}
+\sum'_{\mathbf{q'}}G_{\mathbf{kqq'}}\\
E_{\mathbf{q}}\eta_{\mathbf{q}}&=&\frac{g_0}{U_0}\varphi_{\mathbf{q}}\\
E_{\mathbf{kk'qq'}}4\alpha_{\mathbf{kk'qq'}}&=&2g_0(\Lambda_{\mathbf{k'}}\beta_{\mathbf{kqq'}}-\Lambda_{\mathbf{k}}\beta_{\mathbf{k'qq'}})
+U_0\left[(\phi_{\mathbf{kq}}-\phi_{\mathbf{kq'}})-(\phi_{\mathbf{k'q}}-\phi_{\mathbf{k'q'}})\right]-G_{\mathbf{kqq'}}+G_{\mathbf{k'qq'}}\\
E_{\mathbf{kqq'}}\beta_{\mathbf{kqq'}}&=&-\frac{g_0}{2U_0}G_{\mathbf{kqq'}}\label{p5}
\end{eqnarray}
where $\varphi_{\mathbf{q}}=U_0\sum'_{\mathbf{k}}\phi_{\mathbf{kq}},~G_{\mathbf{kqq'}}=4U_0\sum'_{\mathbf{K}}\alpha_{\mathbf{kKqq'}}$ and we have defined:
\begin{eqnarray}
E_{\mathbf{q}}&=&E+\epsilon^u_{\mathbf{q}}-\epsilon^b_{\mathbf{q}}-\nu_0~~~~~~~~~~~~~~~~~~~~~~~~~~~~~~~~~E_{\mathbf{kq}}=\epsilon_{\mathbf{k}}^u+\epsilon_{\mathbf{q-k}}^d-\epsilon_{\mathbf{q}}^u-E\nonumber\\
E_{\mathbf{kqq'}}&=&E+\epsilon^u_{\mathbf{q}}+\epsilon^u_{\mathbf{q'}}-\epsilon^u_{\mathbf{k}}-\epsilon^b_{\mathbf{q+q'-k}}-\nu_0~~~~~~~~~~~~E_{\mathbf{kk'qq'}}=\epsilon^u_{\mathbf{k}}+\epsilon^u_{\mathbf{k'}}+\epsilon^d_{\mathbf{q+q'-k-k'}}-\epsilon^u_{\mathbf{q}}-\epsilon^u_{\mathbf{q'}}-E\nonumber
\end{eqnarray}
Finally, after eliminating all the variables other than $G_{\mathbf{kqq'}}$ from equation (\ref{p1})-(\ref{p5}) and using the renormalization relations between $U_0,~g_0, ~\nu_0$ and $U_r,~g_r, ~\nu_r$ as given in our paper, we obtain the following integral equations:
\begin{align}
&E=\sum'_{\mathbf{q}}\Gamma^{-1}_2({\bf q},E+\epsilon^\text{u}_{\mathbf{q}})\left(1+\sum^\prime_{\mathbf{k}{\bf q^\prime}}G_{\mathbf{k q q^\prime}}/E_{\mathbf{k q}}\right),\\
&\gamma_{\mathbf{kqq'}}G_{\mathbf{kqq'}}=\phi_{\mathbf{kq'}}-\phi_{\mathbf{kq}}+\sum_{\mathbf{k'}}G_{\mathbf{k'qq'}}/E_{\mathbf{kk'qq'}},\\
&\phi_{\mathbf{kq}}\!=\!\frac{1}{E_{\mathbf{kq}}}\!\left[\sum^\prime\limits_{\mathbf{q^\prime}}G_{\mathbf{kqq'}}\!-\!\Gamma^{-1}_2(q,E\!+\!\epsilon^\text{u}_{\mathbf{q}})\left(1\!+\!\sum^\prime\limits_{\mathbf{k^\prime}{\bf q^\prime}}\frac{G_{\mathbf{k^\prime q q^\prime}}}{E_{\mathbf{k^\prime q}}}\right)\right],
\end{align}
where $\gamma_{\mathbf{kqq'}}=\Gamma^{-1}_2(\mathbf{q+q'-k},E+\epsilon^\text{u}_{\mathbf{q}}+\epsilon^\text{u}_{\mathbf{q'}}-\epsilon^\text{u}_{\mathbf{k}})$, $E_{\mathbf{kq}}=\epsilon_{\mathbf{k}}^\text{u}+\epsilon_{\mathbf{q-k}}^\text{d}-\epsilon_{\mathbf{q}}^\text{u}-E$ and $E_{\mathbf{kk^\prime qq^\prime}}=\epsilon^\text{u}_{\mathbf{k}}+\epsilon^\text{u}_{\mathbf{k^\prime}}+\epsilon^\text{d}_{\mathbf{q+q^\prime-k-k^\prime}}-\epsilon^\text{u}_{\mathbf{q}}-\epsilon^\text{u}_{\mathbf{q^\prime}}-E$. $\Gamma^{-1}_2$ has been defined in equation (6) in our paper. The above integral equations are solved by numeric iterations to obtain the polaron energy $E$.

\subsection{Three-body problem and the atom-dimer scattering length}

In this section of supplementary material we obtain the atom-molecule scattering length by solving the three-body problem using a two-channel model (see equation (1)-(3) in our paper).
We first write he three-body wave function in the following second quantized form in momentum space:
\begin{eqnarray}
|\psi_{3}\rangle=\left[\sum_{\mathbf{k_1 k_2}}\phi_{\mathbf{k_1 k_2}}u^{\dag}_{\mathbf{k_1}}u^{\dag}_{\mathbf{k_2}}d^{\dag}_{\mathbf{-k_1-k_2}}
+\sum_{\mathbf{p}}\eta_{\mathbf{p}}u^{\dag}_{\mathbf{p}}b^{\dag}_{\mathbf{-p}} \right]|\text{vac}\rangle\label{threebody}
\end{eqnarray}
From the Schr\"odinger equation $\hat{H}|\psi_{3}\rangle=E|\psi_{3}\rangle$, we have:
\begin{eqnarray}
&~&(\epsilon_{\mathbf{k}}^u+\epsilon_{\mathbf{k'}}^u+\epsilon_{\mathbf{k+k'}}^d-E)\phi_{\mathbf{k k'}}+U_0\sum_{\mathbf{p}}(\phi_{\mathbf{p k'}}-\phi_{\mathbf{p k}})+\frac{1}{2}g_0(\eta_{\mathbf{k'}}-\eta_{\mathbf{k}})=0\\
&~&(\epsilon_{\mathbf{k}}^u+\epsilon_{\mathbf{k}}^b+\nu_0-E)\eta_{\mathbf{k}}+2g_0\sum_{\mathbf{k'}}\phi_{\mathbf{k'k}}=0
\end{eqnarray}

Defining $\alpha_{\mathbf{k}}=U_0\sum_{\mathbf{k'}}\phi_{\mathbf{k'k}}$ and using the renormalization relation  (here the introducing of the prefactor $U_0$ is just to eliminate the ultra-violate divergence in $\sum_{\mathbf{k'}}$, this divergence corresponds to the $\frac{1}{r}$ singularity in the real space wave function), we have:
\begin{eqnarray}
\gamma_{\mathbf{k}}\alpha_{\mathbf{k}}&=&\sum_{\mathbf{k'}}\frac{\alpha_{\mathbf{k'}}}{E_{\mathbf{k'k}}}\\
\eta_{\mathbf{k}}&=&\frac{2g_r}{U_r}\frac{\alpha_{\mathbf{k}}}{E_{\mathbf{k}}}\label{etak}
\end{eqnarray}
where $E_{\mathbf{k}}=E-\epsilon_{\mathbf{k}}^u-\epsilon_{\mathbf{k}}^b-\nu_0,~E_{\mathbf{k'k}}=\epsilon_{\mathbf{k}}^u+\epsilon_{\mathbf{k'}}^u+\epsilon_{\mathbf{k+k'}}^d-E$ and we define:
\begin{eqnarray}
\gamma_{\mathbf{k}}=\left(U_r+\frac{g_r^2}{E-\epsilon_{\mathbf{k}}^u-\epsilon_{\mathbf{k}}^b-\nu_r} \right)^{-1}+\sum_{\mathbf{k'}}\left(\frac{1}{E_{\mathbf{kk'}}}-\frac{1}{\epsilon^r_{\mathbf{k'}}} \right)
\end{eqnarray}
The fourier transform $\alpha(\mathbf{r})=(2\pi)^3\int d^3\mathbf{k}\alpha_{\mathbf{k}}e^{i\mathbf{k}\cdot\mathbf{r}}$ is just the asymptotic wave function between an atom and a molecule.
To see this, we first write wave function (\ref{threebody}) in real space:
\begin{eqnarray}
|\psi_{3b}\rangle&=&\int d^3\mathbf{\mathbf{r_1}}d^3\mathbf{\mathbf{r_2}}d^3\mathbf{\mathbf{r_3}}\left[\phi_o(\mathbf{r_1},\mathbf{r_2},\mathbf{r_3})u^{\dag}(\mathbf{r_1})u^{\dag}(\mathbf{r_2})d^{\dag}(\mathbf{r_3})
+\phi_c(\mathbf{r_1},\mathbf{r_2},\mathbf{r_3})u^{\dag}(\mathbf{r_1})b^{\dag}(\mathbf{r_3}) \right]|\text{vac}\rangle\label{threebody2}\\
\phi_o(\mathbf{r_1},\mathbf{r_2},\mathbf{r_3})&=&\sum_{\mathbf{k_1 k_2}}\phi_{\mathbf{k_1 k_2}}e^{i\mathbf{k_1}\cdot(\mathbf{r_1}-\mathbf{r_3})}e^{i\mathbf{k_2}\cdot(\mathbf{r_2}-\mathbf{r_3})}\\
\phi_c(\mathbf{r_1},\mathbf{r_2},\mathbf{r_3})&=&\delta^3(\mathbf{r_2}-\mathbf{r_3})\sum_{\mathbf{p}}\eta_{\mathbf{p}}e^{i\mathbf{p}\cdot(\mathbf{r_1}-\mathbf{r_3})}=\delta^3(\mathbf{r_2}-\mathbf{r_3})\eta(\mathbf{r_1}-\mathbf{r_3})
\end{eqnarray}
In the limit $\mathbf{r_2}-\mathbf{r_3}\rightarrow 0$, we have $\phi_o(\mathbf{r_1},\mathbf{r_2},\mathbf{r_3})\simeq\sum_{\mathbf{k_1}}\big(\sum_{\mathbf{k_2}}\phi_{\mathbf{k_1 k_2}}\big)e^{i\mathbf{k_1}\cdot(\mathbf{r_1}-\mathbf{r_3})}=\frac{1}{U_0}\alpha(\mathbf{r_1}-\mathbf{r_3})$ and $\phi_c(\mathbf{r_1},\mathbf{r_2},\mathbf{r_3})=\delta^3(\mathbf{r_2}-\mathbf{r_3})\eta(\mathbf{r_1}-\mathbf{r_3})$. On the other hand, from equation (\ref{etak}) we can see that $\eta_{\mathbf{k}}$ and $\alpha_{\mathbf{k}}$ has identical singularity as $k\rightarrow 0$ (since $\frac{1}{E_{\mathbf{k}}}$ is analytical near $k=0$). This means that in real space, $\eta(\mathbf{r})$ and $\alpha(\mathbf{r})$ have the same long range behavior as $r\rightarrow \infty$. As a result, the long range behavior of $\alpha(\mathbf{r})$ represents the asymptotic wave function between an atom and a molecule as they separate far away from each other.

 As $r\rightarrow \infty$, we have $\alpha(\mathbf{r})=1-\frac{a_{am}}{r}(1+O(r^{-1}))$ where $a_{am}$ is the atom-molecule scattering length. This leads to the fact that
 $\alpha_{\mathbf{k}}=\delta^3(\mathbf{k})-\frac{1}{2\pi^2}\frac{a_{am}}{k^2}(1+O(k))$ as $k\rightarrow 0$. As a result, it will be very convenient to define $\alpha_{\mathbf{k}}=\delta^3(\mathbf{k})-\frac{1}{2\pi^2}\frac{\beta_{\mathbf{k}}}{k^2}$ and we have:
\begin{eqnarray}
\beta_{\mathbf{k}}=\frac{-k^2}{4\pi\gamma_{\mathbf{k}}}\left[\frac{1}{E_{0\mathbf{k}}}-4\pi\int\frac{d^3\mathbf{q}}{8\pi^3}\frac{\beta_{\mathbf{q}}}{q^2E_{\mathbf{qk}}} \right]
\end{eqnarray}
The above integral equation is solved numerically and the atom-molecule scattering length is given as $a_{am}=\beta_{\mathbf{0}}$.

\subsection{Induced Interaction}
In this section of supplementary material, we evaluate the induced interaction $g_{\text{ind}}$ within Born approximation.
The corresponding diagram for the T-matrix of molecule-molecule scattering is shown in Fig. \ref{born}. We have $g_{\text{ind}}=\frac{4\pi\hbar^2 a_{\text{ind}}}{m_{\uparrow}+m_{\downarrow}}=T_m$ and:
\begin{eqnarray}
T_m=g_r^4\int\frac{d\omega}{2\pi}\int\frac{d^3\mathbf{p}}{(2\pi)^3}\frac{1}{(i\omega-\xi_\mathbf{p}^{\uparrow})^2}\frac{1}{(i\omega+\xi_\mathbf{p}^{\downarrow})^2}
\end{eqnarray}
where $g_r^2=\frac{2\pi a_{bg}W}{m_r}$, $\xi_{\mathbf{p}}^{\sigma}=\frac{\mathbf{p}^2}{2m_{\sigma}}-\mu_{\sigma}$ and $\mu_{\downarrow}=E_m$ is given by the molecular energy in the single impurity problem. All the external frequency and momentums are set to zero.

After performing the frequency integral and some change of variables, we finally have:
\begin{align}
k_\text{F}a_{\text{ind}}&=\frac{1}{\pi}\gamma(1+\gamma^{-1})^3\Big(k_Fa_{bg}\frac{W}{\epsilon_F}\Big)^2F(\gamma,\frac{|E_m|}{\epsilon_F})\\
F(\gamma,\eta)&=2\int_1^{\infty}dx\frac{x^2}{[(1+\gamma^{-1})x^2+\eta-1]^3}-\frac{1}{2(\gamma^{-1}+\eta)^2}
\end{align}
where $\gamma=\frac{m_{\downarrow}}{m_{\uparrow}}$.

\begin{figure}[tbp]
\includegraphics[height=2.3in, width=2.1in]
{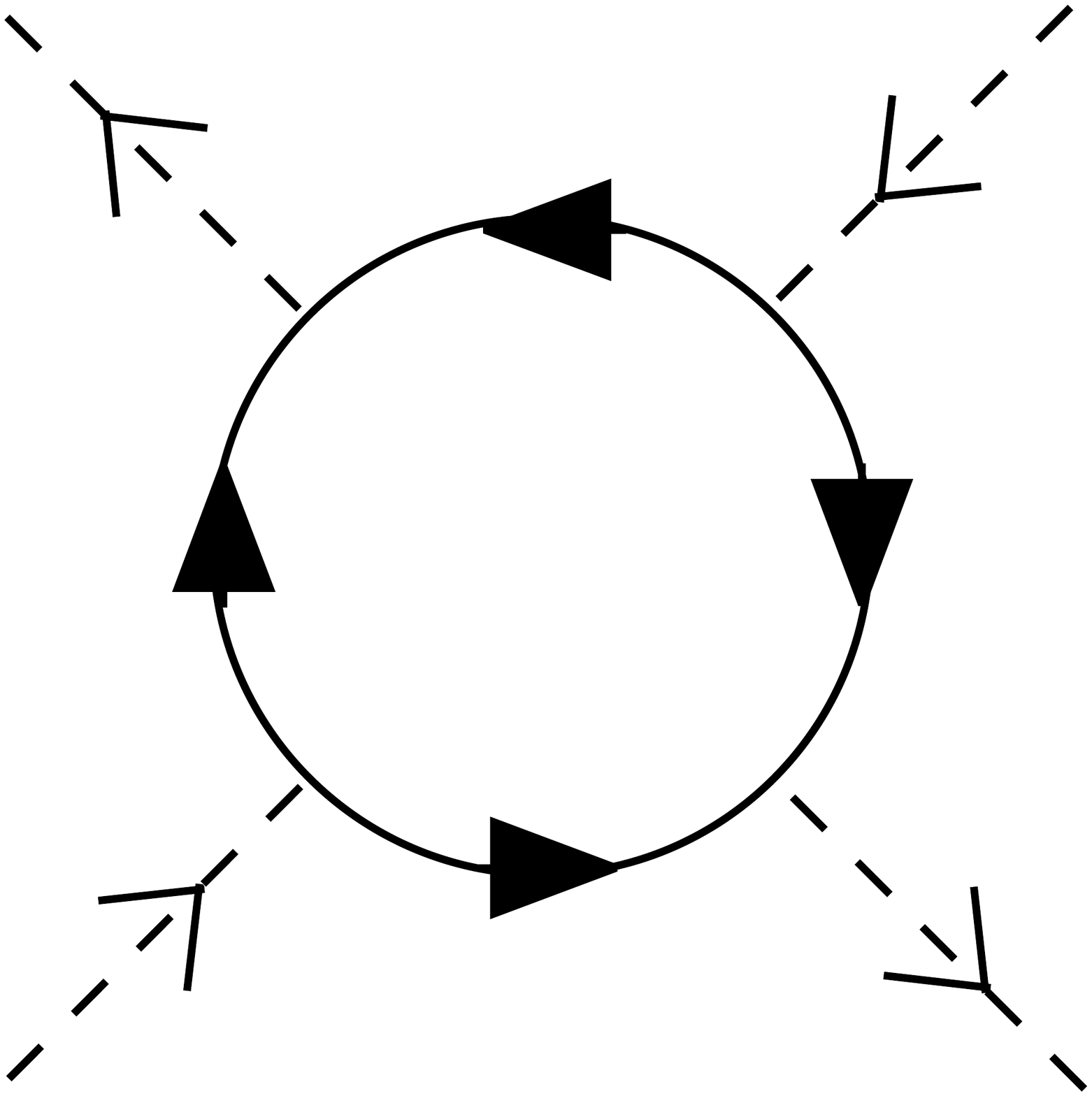}
\caption{Induced molecule-molecule interaction under Born approximation. The dashed and solid line refers to the propagator of molecule and atoms respectively. Each vertex is given by the inter-channel coupling $g_r$.}\label{born}
\end{figure}

\end{widetext}


\begin{thebibliography}{99}
\bibitem{residue}
In our variational approach, the quasi-particle residue is just given as $Z=|\phi_0|^2$ with $\phi_0$ solved from (5).
\bibitem{stability}
This expansion of energy functional is only valid for sufficiently narrow resonance where closed channel molecule is dominant in the pairing state.
\end{thebibliography}
\end{document}